\documentstyle[11pt,amssymb,epsfig,psfig]{article}

\begin{document}

\title{{\bf Stress- Energy Tensor for Parallel Plate on Background of
 Conformally Flat Brane-World Geometries and Cosmological Constant Problem }}
\author{M. R. Setare  \footnote{E-mail: rezakord@ipm.ir }
 \\
 {Physics Dept. Inst. for Studies in Theo. Physics and
Mathematics(IPM)}\\
{P. O. Box 19395-5531, Tehran, IRAN } }

\date{\small{\today}}

\maketitle

\begin{abstract}
In this paper, we calculate the stress-energy tensor for a
quantized massless conformally coupled scalar field in the
background of a conformally flat brane-world geometries, where the
scalar field satisfying Robin boundary conditions on two parallel
plates. In the general case of Robin boundary conditions formula
are derived for the vacuum expectation values of the
energy-momentum tensor. Further the surface energy per unit area
are obtained . As an application of the general formula we have
considered the important special case of the AdS$_{4+1}$ bulk,
moreover application to the Randall-Sundrum scenario is discused.
In this specific example for a certain choice of Robin
coefficients, one could make the effective cosmological constant
vanish.
\end{abstract}

\newpage

\section{Introduction}
The cosmological constant was first introduced by Einstein in
order to justify the equilibrium of a static universe against its
own gravitational attraction. The discovery of Hubble that the
universe may be expanding led Einstein to abandon the idea of a
static universe and, along with it the cosmological constant.
However the Einstein static universe remained to be of interest to
theoreticians since it provided a useful model to achieve better
understanding of the interplay of spacetime curvature and of
quantum field theoretic effects. Recent year have witnessed a
resurgence of interest in the possibility that a positive
cosmological constant $\Lambda $ may dominate the total energy
density in the universe  \cite{{9},{Carrol},{Sahni}}. At a
theoretical level $\Lambda $ is predicted to arise out of the
zero-point quantum vacuum fluctuations of the fundamental quantum
fields. Using parameters arising in the electroweak theory results
in a value of the vacuum energy density $\rho _{vac}=10^{6}$
GeV$^{4}$ which is almost $10^{53}$ times larger than the current
observational upper limit on $\Lambda $ which is $10^{-47}$
GeV$^{4}\sim 10^{-29}$ gm/cm$^{3}$. On the other hand the QCD
vacuum is expected to generate a cosmological constant of the
order of $10^{-3}$ GeV$^{4}$ which is many orders of magnitude
larger than the observed value. This is known as the old
cosmological constant problem. The new cosmological problem is to
understand why $\rho _{vac}$ is not only small but also, as the
current observations seem to indicate, is of the same order of
magnitude as the present mass density of the universe.\\
In recent years, there has been a hope to understand the vanishing
cosmological constant in extera dimensional theories
\cite{kach}-\cite{kim2}. It is generally believed that fine-tuning
is necessary for a very small cosmological constant in
4-dimensional theories \cite{{fla},{bin},{for}}. This leads one to
search for a naturally small cosmological constant in higher
dimensional theories. However, for a usual compactification of a
higher dimensional theory to an effective 4-dimensional theory,
one ends up with a normal 4-dimensional theory, and the
fine-tuning problem generically reappears. This is the case for
the usual Kaluza-Klein (KK) compactification, and for the generic
compactification with large extra dimension \cite{Rand99}. The
Randall-Sundrum (RS) model \cite{Rand99} provides a hope of
avoiding this pathology. This higher dimensional scenario is based
on a non-factorizable geometry which accounts for the ratio
between the Planck scale and weak scales without the need to
introduce a large hierarchy between fundamental Planck scale and
the compactification scale. The model consists of a spacetime with a single $%
S^1/Z_2$ orbifold extra dimension. Three-branes with opposite
tension reside at the orbifold fixed points, and together with a
finely tuned negative bulk cosmological constant serve as sources
for five-dimensional gravity.\\
 In the present paper we will
investigate the vacuum expectation values of the energy-momentum
tensor of the conformally coupled scalar field on background of
the conformally flat Brane-World geometries. We will consider the
general plane--symmetric solutions of the gravitational field
equations and boundary conditions of the Robin type on the branes.
The latter includes the Dirichlet and Neumann boundary conditions
as special cases. The Casimir energy-momentum tensor for these
geometries can be generated from the corresponding flat spacetime
results by using the standard transformation
formula\cite{{Seta01b},{setsah}}. Previously this method has been
used in \cite{Seta01b} to derive the vacuum stress on parallel
plates for a scalar field with Dirichlet boundary conditions in
de Sitter spactime and in Ref. \cite{setsah}to investigate the
vacuum characteristics of the Casimir configuration on background
of conformally flat brane-world geometries for massless scalar
field with Robin boundary conditions on plates.
 Also this method has been used in \cite{Seta01} to
derive the vacuum characteristics of the Casimir configuration on
background of the static domain wall geometry for a scalar field
with Dirichlet boundary condition on plates.(for investigations
of the Casimir energy in braneworld models with dS branes see
    Refs. \cite{fab,gariga1, shin, gariga2,klem, nay}). For Neumann or more
general mixed boundary conditions we need to have the Casimir
energy-momentum tensor for the flat spacetime counterpart in the
case of the Robin boundary conditions with coefficients related to
the metric components of the brane-world geometry and the boundary
mass terms. The Casimir effect for the general Robin boundary
conditions on background of the Minkowski spacetime was
investigated in Ref. \cite{RomSah} for flat boundaries, and in
\cite{Saha01a,Saha01b} for spherically and cylindrically symmetric
boundaries in the case of a general conformal coupling( For
Robin-type condition see also\cite{vas1,vas2})\footnote{Further
developments in Csimir effect can be found in \cite{fuling}.} .
Here we use the results of Ref. \cite{RomSah} to generate vacuum
energy--momentum tensor for the plane symmetric conformally flat
backgrounds, in second section we review this work briefly,
further in section $3$ the surface energy per unit area which
located on the branes, are obtained. This surface term is zero for
Dirichlet or Neumann boundary condition but yields a nonvanishing
contribution for Robin boundary conditions. In the general case
(general coupling), the stress energy tensor diverges close to the
branes. This would also be expected in the conformal case if the
branes are curved \cite{deu}. In section ${4}$ the important
special case of $AdS$ background is considered, we obtain an
explicit relation between the cosmological constant of AdS$_{4+1}$
bulk and brane tension (which is the surface energy per unit area
where located on the branes), then the application to the
Randall-Sundrum is discussed. Finally, the results are
re-mentioned and discussed in last section.

\section{Vacuum Expectation values for the Energy-Momentum Tensor}
In this paper we will consider a conformally coupled massless scalar field $%
\varphi (x)$ satisfying the equation
\begin{equation}
\left( \nabla _{\mu }\nabla ^{\mu }+\xi R\right) \varphi
(x)=0,\quad \xi =\frac{D-1}{4D}  \label{fieldeq}
\end{equation}
on background of a $D+1$--dimensional conformally flat
plane--symmetric spacetime with the metric
\begin{equation}
g_{\mu \nu }=e^{-2\sigma (z)}\eta _{\mu \nu },\quad \mu ,\nu
=0,1,\ldots ,D. \label{metric}
\end{equation}
In Eq. (\ref{fieldeq}) $\nabla _{\mu }$ is the operator of the
covariant derivative, and $R$ is the Ricci scalar for the metric
$g_{\mu \nu }$. Note that for the metric tensor from Eq.
(\ref{metric}) one has
\begin{equation}
R=De^{2\sigma }\left[ 2\sigma ^{\prime \prime }-(D-1)\sigma ^{\prime 2}%
\right] ,  \label{Riccisc}
\end{equation}
where the prime corresponds to the differentiation with respect to
$z$.

We will assume that the field satisfies the mixed boundary
condition
\begin{equation}
\left( a_{j}+b_{j}n^{\mu }\nabla _{\mu }\right) \varphi
(x)=0,\quad z=z_{j},\quad j=1,2  \label{boundcond}
\end{equation}
on the hypersurfaces $z=z_{1}$ and $z=z_{2}$, $z_{1}<z_{2}$,
$n^{\mu }$ is the normal to these surfaces, $n_{\mu }n^{\mu }=-1$,
and $a_j$, $b_j$ are constants. The results in the following will
depend on the ratio of these coefficients only. However, to keep
the transition to the Dirichlet and Neumann cases transparent we
will use the form (\ref{boundcond}). For the case of plane
boundaries under consideration introducing a new coordinate $y$ in
accordance with
\begin{equation}
dy=e^{-\sigma }dz  \label{ycoord}
\end{equation}
conditions (\ref{boundcond}) take the form
\begin{equation}
\left( a_{j}+(-1)^{j-1}b_{j}e^{\sigma (z_{j})}\partial _{z}\right)
\varphi (x)=\left( a_{j}+(-1)^{j-1}b_{j}\partial _{y}\right)
\varphi (x)=0,\quad y=y_{j},\quad j=1,2.  \label{boundcond1}
\end{equation}
Note that the Dirichlet and Neumann boundary conditions are
obtained from Eq. (\ref{boundcond}) as special cases corresponding
to $(a_j,b_j)=(1,0)$ and $(a_j,b_j)=(0,1)$ respectively. Our main
interest in the present paper is to investigate the vacuum
expectation values (VEV's) of the energy--momentum tensor for the field $%
\varphi (x)$ in the region $z_{1}<z<z_{2}$. The presence of
boundaries modifies the spectrum of the zero--point fluctuations
compared to the case without boundaries. This results in the shift
in the VEV's of the physical quantities, such as vacuum energy
density and stresses. This is the well known Casimir effect.

It can be shown that for a conformally coupled scalar by using
field equation (\ref{fieldeq}) the expression for the
energy--momentum tensor can be presented in the form
\cite{Birrell}
\begin{equation}
T_{\mu \nu }=\nabla _{\mu }\varphi \nabla _{\nu }\varphi -\xi \left[ \frac{%
g_{\mu \nu }}{D-1}\nabla _{\rho }\nabla ^{\rho }+\nabla _{\mu
}\nabla _{\nu }+R_{\mu \nu }\right] \varphi ^{2},  \label{EMT1}
\end{equation}
where $R_{\mu \nu }$ is the Ricci tensor. The quantization of a
scalar filed on background of metric (2) is standard. Let
$\{\varphi _{\alpha }(x),\varphi _{\alpha }^{\ast }(x)\}$ be a
complete set of orthonormalized positive and negative frequency
solutions to the field equation (\ref {fieldeq}), obying boundary
condition (\ref{boundcond}). By expanding the field operator over
these eigenfunctions, using the standard commutation rules and the
definition of the vacuum state for the vacuum expectation values
of the energy-momentum tensor one obtains
\begin{equation}
\langle 0|T_{\mu \nu }(x)|0\rangle =\sum_{\alpha }T_{\mu \nu }\{\varphi {%
_{\alpha },\varphi _{\alpha }^{\ast }\}},  \label{emtvev1}
\end{equation}
where $|0\rangle $ is the amplitude for the corresponding vacuum
state, and the bilinear form $T_{\mu \nu }\{{\varphi ,\psi \}}$ on
the right is determined by the classical energy-momentum tensor
(\ref{EMT1}). In the problem under consideration we have a
conformally trivial situation: conformally invariant field on
background of the conformally flat spacetime. Instead of
evaluating Eq. (\ref{emtvev1}) directly on background of the
curved metric, the vacuum expectation values can be obtained from
the corresponding flat spacetime results for a scalar field
$\bar{\varphi}$ by using the conformal properties of the problem
under consideration. Under the
conformal transformation $g_{\mu \nu }=\Omega ^{2}\eta _{\mu \nu }$ the $%
\bar{\varphi}$ field will change by the rule
\begin{equation}
\varphi (x)=\Omega ^{(1-D)/2}\bar{\varphi}(x),  \label{phicontr}
\end{equation}
where for metric (\ref{metric}) the conformal factor is given by
$\Omega =e^{-\sigma (z)}$. The boundary conditions for the field
$\bar{\varphi}(x)$ we will write in form similar to Eq.
(\ref{boundcond1})
\begin{equation}
\left( \bar{a}_{j}+(-1)^{j-1}\bar{b}_{j}\partial _{z}\right) \bar{\varphi}%
=0,\quad z=z_{j},\quad j=1,2,  \label{bounconflat}
\end{equation}
with constant Robin coefficients $\bar{a}_{j}$ and $\bar{b}_{j}$.
Comparing to the boundary conditions (\ref{boundcond}) and taking
into account transformation rule (\ref{phicontr}) we obtain the
following relations between the corresponding Robin coefficients
\begin{equation}
\bar{a}_{j}=a_{j}+(-1)^{j-1}\frac{D-1}{2}\sigma ^{\prime
}(z_{j})e^{\sigma (z_{j})}b_{j},\quad \bar{b}_{j}=b_{j}e^{\sigma
(z_{j})}.  \label{coefrel}
\end{equation}
Note that as Dirichlet boundary conditions are conformally
invariant the Dirichlet scalar in the curved bulk corresponds to
the Dirichlet scalar in a flat spacetime. However, for the case of
Neumann scalar the flat spacetime counterpart is a Robin scalar
with $\bar{a}_j=(-1)^{j-1}(D-1)\sigma '(z_j)/2$ and $\bar{b}_j=1$.
The Casimir effect with boundary conditions (\ref{bounconflat}) on
two parallel plates on background of the Minkowski spacetime is
investigated in Ref. \cite{RomSah} for a scalar field with a
general conformal coupling parameter. In the case of a conformally
coupled scalar the corresponding regularized VEV's for the
energy-momentum tensor are uniform in the region between the
plates and have the form
\begin{equation}
\langle \bar{T}_{\nu }^{\mu }\left[ \eta _{\alpha \beta }\right] \rangle _{%
{\rm ren}}=-\frac{J_D(B_1,B_2)}{2^{D}\pi ^{D/2}a^{D+1}\Gamma
(D/2+1)}{\rm diag}(1,1,\ldots ,1,-D), \quad z_{1}< z< z_{2},
\label{emtvevflat}
\end{equation}
where
\begin{equation}\label{IDB1B2}
  J_D(B_1,B_2)={\rm p.v.}
\int_{0}^{\infty }\frac{t^{D}dt}{\frac{(B_{1}t-1)(B_{2}t-
1)}{(B_{1}t+1)(B_{2}t+1)}e^{2t}-1},
\end{equation}
and we use the notations
\begin{equation}
B_{j}=\frac{\bar{b}_{j}}{\bar{a}_{j}a},\quad j=1,2,\quad
a=z_{2}-z_{1}. \label{Bjcoef}
\end{equation}
For the Dirichlet and Neumann scalars $B_1=B_2=0$ and
$B_1=B_2=\infty $ respectively, and one has
\begin{equation}\label{JDDirNeu}
  J_D(0,0)=J_D(\infty ,\infty )=\frac{\Gamma (D+1)}{2^{D+1}}\zeta
  _R(D+1),
\end{equation}
with the Riemann zeta function $\zeta _R(s)$. Note that in the
regions $z< z_{1}$ and $z> z_{2}$ the Casimir densities vanish
\cite{RomSah}:
\begin{equation}
\langle \bar{T}_{\nu }^{\mu }\left[ \eta _{\alpha \beta }\right] \rangle _{%
{\rm ren}}=0,\quad z< z_{1},z> z_{2}.  \label{emtvevflat2}
\end{equation}
This can be also obtained directly from Eq. (\ref{emtvevflat})
taking the limits $z_{1}\rightarrow -\infty $ or $z_{2}\rightarrow
+\infty $.\\
The vacuum energy-momentum tensor on curved background
(\ref{metric}) is obtained by the standard transformation law
between conformally related problems (see, for instance,
\cite{Birrell}) and has the form
\begin{equation}
\langle T_{\nu }^{\mu }\left[ g_{\alpha \beta }\right] \rangle _{{\rm ren}%
}=\langle T_{\nu }^{\mu }\left[ g_{\alpha \beta }\right] \rangle _{{\rm ren}%
}^{(0)}+\langle T_{\nu }^{\mu }\left[ g_{\alpha \beta }\right] \rangle _{%
{\rm ren}}^{(b)}.  \label{emtcurved1}
\end{equation}
Here the first term on the right is the vacuum energy--momentum
tensor for the situation without boundaries (gravitational part),
and the second one is due to the presence of boundaries. As the
quantum field is conformally coupled and the background spacetime
is conformally flat the gravitational part of the energy--momentum
tensor is completely determined by the trace anomaly and is
related to the divergent part of the corresponding effective
action by the relation \cite{Birrell}
\begin{equation}
\langle T_{\nu }^{\mu }\left[ g_{\alpha \beta }\right] \rangle _{{\rm ren}%
}^{(0)}=2g^{\mu \sigma }(x)\frac{\delta }{\delta g^{\nu \sigma }(x)}W_{{\rm %
div}}[g_{\alpha \beta }].  \label{gravemt}
\end{equation}
Note that in odd spacetime dimensions the conformal anomaly is
absent and the corresponding gravitational part vanishes:
\begin{equation}
\langle T_{\nu }^{\mu }\left[ g_{\alpha \beta }\right] \rangle _{{\rm ren}%
}^{(0)}=0,\quad {\rm for\;even}\;D.  \label{gravemteven}
\end{equation}
The boundary part in Eq. (\ref{emtcurved1}) is related to the
corresponding flat spacetime counterpart
(\ref{emtvevflat}),(\ref{emtvevflat2}) by the relation
\cite{Birrell}
\begin{equation}
\langle T_{\nu }^{\mu }\left[ g_{\alpha \beta }\right] \rangle _{{\rm ren}%
}^{(b)}=\frac{1}{\sqrt{|g|}}\langle \bar{T}_{\nu }^{\mu }\left[
\eta _{\alpha \beta }\right] \rangle _{{\rm ren}}.
\label{translaw}
\end{equation}
By taking into account Eq. (\ref{emtvevflat}) from here we obtain
\begin{equation}
\langle T_{\nu }^{\mu }\left[ g_{\alpha \beta }\right] \rangle _{{\rm ren}%
}^{(b)}=-\frac{e^{(D+1)\sigma (z)}J_D(B_1,B_2)}{2^{D}\pi ^{D/2}a^{D+1}\Gamma (D/2+1)}%
{\rm diag}(1,1,\ldots ,1,-D),  \label{bpartemt}
\end{equation}
for $z_{1}< z< z_{2}$, and
\begin{equation}
\langle T_{\nu }^{\mu }\left[ g_{\alpha \beta }\right] \rangle _{{\rm ren}%
}^{(b)}=0,\;{\rm for}\;z< z_{1},z> z_{2}.  \label{bpartemt2}
\end{equation}
In Eq. (\ref{bpartemt}) the constants $B_{j}$ are related to the
Robin coefficients in boundary condition (\ref{boundcond}) by
formulae (\ref {Bjcoef}),(\ref{coefrel}) and are functions on
$z_j$. In particular, for Neumann boundary conditions
$B^{(N)}_j=2(-1)^{j-1}/[a(D-1)\sigma '(z_j)]$.
\section{Surface Energy Tensor and Branes Tension}
The total bulk vacuum energy per unit physical hypersurface on the
brane at $z=z_j$ is obtained by integrating over the region
between the plates
\begin{equation}\label{bulktoten}
  E_j^{(b)}=e^{D\sigma (z_j)}\int _{z_1}^{z_2}\langle T_0^0\rangle
  ^{(b)}_{{\mathrm{ren}}}e^{-(D+1)\sigma (z)}dz=-\frac{J_D(B_1,B_2)
  e^{D\sigma (z_j)}}{2^D\pi ^{D/2}\Gamma (D/2+1)a^D},
\end{equation}
this result differs from the total Casimir energy per unit volume,
the reason for this difference should be the existence of an
additional surface energy contribution to the volume energy.  The
corresponding energy density is defined by the relation (see,
\cite{RomSah})
\begin{equation}
T_{00}^{{\rm (surf)}}=-\frac{4\xi -1}{2}\delta (z;\partial
M)\varphi
\partial _{z}\varphi ,  \label{surfen1pl}
\end{equation}
located on the boundaries $z=z_{j}$, $j=1,2$,
 where now
\begin{equation}
\quad \delta (z;\partial M)=\delta(z-z_{2}-0) -\delta (z-z_{1}+0),
\label{surfen1}
\end{equation}
where $\delta(z-z_{j}\pm0)$ is a one sided $\delta-$distribution.
In the general case (general coupling), the stress energy tensor
diverges close to the branes. This would also be expected in the
conformal case if the branes are curved \cite{deu}. But in our
case from the above formula it follows that the surface term is
zero for
Dirichlet or Neumann boundary condition (as the factors $\varphi $ or $%
\partial _{z}\varphi $ would then vanish) but yields a nonvanishing
contribution for Robin boundary conditions. The corresponding
v.e.v. can be evaluated by the standard method explained in the
\cite{RomSah}. This leads to the formula
\begin{equation}
\begin{array}{lll}
\langle 0\left| T_{00}^{{\rm (surf)}}\right| 0\rangle  & = & \displaystyle%
\frac{4\xi -1}{2}\delta (z;\partial M)\left( \partial _{z}\langle
0\left| \varphi (z)\varphi (z^{\prime })\right| 0\rangle \right)
\mid _{z^{\prime }=z}
\end{array}
\label{surfendens}
\end{equation}
which provides the energy density on the plates themselves. The
integrated surface energy per unit area are given by
\begin{equation}
\begin{array}{rcl}
\varepsilon _{c}^{{\rm (surf)}} & = & \displaystyle\frac{1}{a}%
\int_{z_{1}}^{z_{2}}dz\,\langle 0\left| T_{00}^{{\rm
(surf)}}\right| 0\rangle,
\end{array}
\label{surfen}
\end{equation}
where $a=z_2-z_1$.
 After regularization for the surface energy per unit area one obtains
\begin{equation}
\bar{\rm {E}}^{{\rm (surf)}}=a\varepsilon _{c}^{{\rm (surf)}}=\sum _{j=1}^{2}E^{{\rm (s)(surf)}%
}(\beta _j)-aD(4\xi -1)\varepsilon _{c}^{(2)}  \label{surfen3}
\end{equation}
with $\varepsilon _{c}^{(2)}$ defined as following introduced
notation
\begin{equation}
\varepsilon _{c}^{(2)}={\frac{B_{1}+B_{2}}{2^{D}\pi
^{D/2}a^{D+1}\Gamma
\left( 1+{\frac{D}{2}}\right) }}\,{\rm p.v.}\int_{0}^{\infty }dt\,{\frac{%
t^{D}(1-B_{1}B_{2}t^{2})}{(1-B_{1}t)^{2}(1-B_{2}t)^{2}%
\,e^{2t}-(1-B_{1}^{2}t^{2})(1-B_{2}^{2}t^{2})}.}  \label{epsc2}
\end{equation}
For Dirichlet ($B_{1}=B_{2}=0$) and Neumann ($B_{1}=B_{2}=\infty
$) scalars this term vanishes.  Note that, as it follows from
(\ref{surfen3}), the quantity $\varepsilon _{c}^{(2)}$ is the
additional (to the single plate) surface energy per unit volume in
the case of the conformally coupled scalar field.\\
As follows from Eq.(\ref{surfen}), in our conformally curved
background the surface energy per unit area which located on the
branes are given by
\begin{equation}
E_{j}^{{\rm (surf)}}= e^{D\sigma(z_{j})} \bar{E} ^{{\rm (surf)}}
  \label{cursurfen3}
\end{equation}
As one can see from Eq.(\ref{surfen3}) the vacuum energy per unit
hypersurface on the brane $z=z_j$ can contain terms in the form
$\sum _{j=1}^{2} E^{{\rm (s)(surf)}}(\beta _j)$ with constants
$\beta _1$ and $\beta _2 $ and corresponding to the single brane
contributions when the second brane is absent. Adding these terms
to the vacuum energy corresponds to finite renormalization of the
tension on both branes.
\section{Casimir Surface Energy  on the Branes in  AdS$_{4+1}$ Bulk and Cosmological Constant Problem}
\label{sec:dS} As an application of the general formulae from the
previous section here we consider the important special case of
the AdS$_{4+1}$ bulk for which
\begin{equation}\label{sigads}
  \sigma =\ln (k_4z)=k_4y,
\end{equation}
with $1/k_4$ being the AdS curvature radius. AdS space is a
spacetime that has the maximal symmetry and a negative constant
curvature, supported by a negative cosmological constant. For an
$4+1$-dimensional AdS space, its curvature radius is related to
the cosmological constant by
\begin{equation}\label{cosrad}
k_4=(\frac{-\Lambda}{6}) ^{1/2}
\end{equation}
 Now the expressions for the
coefficients $B_j$, $j=1,2$ take the form
\begin{equation}
B_{j}=\frac{b_{j}k_4z_j}{(z_2-z_1)\left[a_{j}+3(-1)^{j-1}k_{4}b_{j}/2\right]
}. \label{Bjads}
\end{equation}
Note that the ratio $z_2/z_1$ is related to the proper distance
between the branes $\Delta y$ by the formula
\begin{equation}\label{distads}
  z_2/z_1=e^{k_4\Delta y},\quad \Delta y=y_2-y_1.
\end{equation}
For the surface energy per unit area which located on the branes
on has
\begin{equation}
E_{j}^{{\rm (surf)}}= (k_4 z_j) ^{4} \bar{E} ^{{\rm (surf)}}
  \label{cursurfen4}
\end{equation}
Then using Eqs.(\ref{surfen3}, \ref{epsc2},\ref{cursurfen3}) the
surface energy per unit area of branes in the AdS$_{4+1}$ bulk are
given by
\begin{eqnarray}
{\rm {E}}^{{\rm (surf)}}&=& \frac{\Lambda^{2}z_{j}^{4}}{36}(\sum _{j=1}^{2}E^{{\rm (s)(surf)}%
}(\beta _j)\nonumber\\
 &+&{\frac{B_{1}+B_{2}}{16\pi
^{2}a^{4}\Gamma
\left( 3\right) }}\,{\rm p.v.}\int_{0}^{\infty }dt\,{\frac{%
t^{4}(1-B_{1}B_{2}t^{2})}{(1-B_{1}t)^{2}(1-B_{2}t)^{2}%
\,e^{2t}-(1-B_{1}^{2}t^{2})(1-B_{2}^{2}t^{2})}).}
   \label{epsc21}
\end{eqnarray}
 For a two 3-brane with brane tension
$\sigma_{0}$, the effective 4-dimensional cosmological constant as
seen by observer on the brane is taken to be zero, in the other
terms for a certain choice of Robin coefficients, one could make
this vanish,
\begin{equation}\label{concos}
\Lambda_{eff}=\sigma_{0}+E^{(surf)}_{(our brane)
}(\beta)-\sqrt{\frac{6\Lambda^{2}}{\kappa ^{2}}}=0,
\end{equation}
where $\kappa ^{2}$ is the 5-dimensional gravitational coupling,
and $\Lambda$ is the bulk cosmological constant. However,
requiring (\ref{concos}) to cancel is still a fine-tuning.
 Then in our model the boundary condition is
another possibility to make the cosmological constant vanish. We
could obtain this result only in our interesting case (massless
conformally case with general Robin boundary condition in
odd-dimensional spacetimes).\\
Now we turn to the brane--world model introduced by Randall and
Sundrum \cite{Rand99} and based on the AdS geometry with one extra
dimension. The fifth dimension $y$ is compactified on an orbifold,
$S^1/Z_2$ of length $\Delta y$, with $-\Delta y\leq y\leq \Delta
y$. The orbifold fixed points at $y=0$ and $y=\Delta y$ are the
locations of two 3-branes. For the conformal factor in this model
one has $\sigma =k_4|y|$. The boundary conditions for the
corresponding conformally coupled bulk scalars have the form
(\ref{boundcond1}) with Robin coefficients $a_j/b_j=-c_jk_4$,
where the constants $c_j$ are the coefficients in the boundary
mass term \cite{Gher00}:
\begin{equation}\label{boundmass}
  m_{\varphi }^{(b)2}=2k_4\left[ c_1\delta (y)+c_2\delta (y-\Delta
  y)\right] .
\end{equation}
Note that here we consider the general case when the boundary
masses are different for different branes. Supersymmetry requires
$c_2=-c_1$. The surface energy per unit area on the branes in the
Randall-Sundrum brane-world background are obtained from Eq.
(\ref{epsc21}) with additional factor 1/2. This factor is related
to the fact that now in the normalization condition for the
eigenfunctions the integration goes over the region $(-\Delta
y,\Delta y)$, instead of $(0,\Delta y )$. The coefficients $B_j$
in the expression for $J_4(B_1,B_2)$ are given by the formula
\begin{equation}\label{BjRS}
  B_j=-\frac{e^{(j-1)k_4\Delta y}}{e^{k_4\Delta y}-1}\frac{1}{c_j
  +(-1)^{j}3/2}.
\end{equation}
Recently the energy-momentum tensor in the Randall-Sundrum
braneworld for a bulk scalar with zero mass terms $c_{1}$ and
$c_{2}$ is considered in \cite{tom}, see also \cite{sahari}.
\section{Conclusion}
\label{sec:conclusion}

The Casimir effect on two parallel plates in conformally flat
brane-world geometries background due to conformally coupled
massless scalar field satisfying Robin boundary conditions on the
plates is investigated. In the general case of Robin boundary
conditions formulae are derived for the vacuum expectation values
of the energy-momentum tensor from  the corresponding flat
spacetime results by using the conformal properties of the
problem. The purely gravitational part arises due to the trace
anomaly and is zero for odd spacetime dimensions. In the region
between the branes the boundary induced part for the vacuum
energy-momentum tensor is given by formula (\ref{bpartemt}), and
the corresponding total bulk vacuum energy per unit hypersurface
on the brane have the form Eq.(\ref{bulktoten}). Further the
surface energy per unit area which located on the branes  are
given by Eq.(\ref {cursurfen3}). As an application of the general
formula we have considered the important special case of the
AdS$_{4+1}$ bulk. In this specific example we can write the
effective cosmological constant as Eq.(\ref{concos}), for a
certain choice of Robin coefficients, one could make the
effective cosmological constant vanish. However, requiring
Eq.(\ref{concos}) to cancel is still a fine-tuning.
 The surface energy is zero for
Dirichlet or Neuman boundary condition but yields a non vanishing
contribution for Robin boundary conditions. Moreover, there is a
region in the space of Robin parameters in which the interaction
forces between two 3-brane are repulsive for small distances and
are attractive for large distances \cite{{setsah},{sahari}}. This
provides a possibility to stabilize interplate distance by using
the vacuum forces. Then may be one can say that this kind of
boundary condition is more natural for cosmology. On the other
hand, one can think of many quantum effects that contribute
similarly to the brane tension, the Casimir energy from fields
confined on the brane, or the Casimir effect from other type of
bulk field, which might play a role in realistic models.
 An application to the Randall-Sundrum brane-world model is
discussed. In this model the coefficients in the Robin boundary
conditions on branes are related to the boundary mass terms for
the scalar field under consideration.

\end{document}